\documentclass{elsart}
\usepackage{amssymb}
\usepackage[english]{babel}
\usepackage{graphicx}
\usepackage{textcomp}
\usepackage{amsmath}

\numberwithin{equation}{section}

\begin{document}

\begin{frontmatter}

\title{Typhoon eye trajectory based on a mathematical model: comparing with observational data}

\author[1]{Olga S. Rozanova\corauthref{cor1} }
\ead{rozanova@mech.math.msu.su} \corauth[cor1]{Corresponding author}
\author[2]{Jui-Ling Yu}
\author[3]{Chin-Kun Hu}

\address[1]{Department of Mechanics and Mathematrics, Moscow
State University, Moscow 119992 Russia}
\address[2]{Department of Applied Mathematics, Providence
University, Taichung, 43301, Taiwan; and National Center of
Theoretical Sciences at Taipei, Physics Division, National Taiwan
University, Taipei 10617, Taiwan}
\address[3]{Institute of Physics, Academia Sinica, Nankang, Taipei
11529, Taiwan\\ and Center for Nonlinear and Complex Systems and
Department of Physics, Chung Yuan Christian University, Chungli
32023, Taiwan}
\date{\today}

\begin{abstract}
We propose a model based on the primitive system of the
Navier-Stokes equations in a bidimensional framework as the $l$ -
plane approximation, which allows us to explain the variety of
tracks of tropical cyclones (typhoons). Our idea is to construct
special analytical solutions with a linear velocity profile for the
Navier-Stokes systems. The evidence of the structure of linear
velocity near the center of vortex can be proven by the
observational data. We study solutions with the linear-velocity
property for both barotropic and baroclinic cases and show that they
follow the same equations in describing the trajectories of the
typhoon eye at the equilibrium state (that relates to the
conservative phase of the typhoon dynamics). Moreover, at the
equilibrium state, the trajectories can be viewed as a superposition
of two circular motions: one has period $2\pi/l,$ the other one has
period $2\pi/b_0,$ where $l$ is the Coriolis parameter and $b_0$ is
the height-averaged vorticity at the center of cyclone.

Also, we compare our theoretical trajectories based on initial
conditions from the flow with tracks obtained from the observational
database. It is worth to mention that under certain conditions our
results are still compatible with observational data although we did
not truly consider the influence of steering effect.
Finally, we propose the parameter-adopting method so that
one could correct the weather prediction in real time. Examples of
our analysis and the use of parameter-adopting method for the
historic trajectories are provided.
\end{abstract}

\end{frontmatter}


\medskip

\section{Introduction}

Tropical cyclones, commonly known as hurricanes in the North
Atlantic Ocean and typhoons in the western North Pacific Ocean, are
one of the most devastating weather phenomena in the world. The
intense winds associated with tropical cyclone often generate ocean
waves and heavy rain, which usually result in severe disasters. To
reduce the harm caused by the tropical cyclones, the most important
issue is the prediction of its moving trajectory.

For a long time, the phenomenon of formation of a huge intense
atmospheric vortex attracted attentions of scientists working in
various fields (e.g. see \cite{Gray} and references in \cite{Rrcd}).
A large progress has been made in the last decade in understanding
the physical process and the quality of operational prediction of
tropical cyclones (\cite{chan},\cite{Weber}, and references
therein). In particular, statistical and numerical weather
prediction models have been improved dramatically and used to
provide operational forecasters the high-quality prediction for
national weather bureaus.

The system of equations served for the numerical prediction is very
complicated. It needs to take many factors into account, for
examples, phase transitions, vertical advection, and boundary layer
effects, etc. Generally, it is difficult to use such system to
comprehend its underlying physical processes. To have a conceptual
understanding about these processes, some assumptions must be made
to simplify the equations. The solutions then may appear in a
simpler form so that it becomes much easier to gain their physical
interpretations. However, neglecting some terms in the equations may
risk losing some important properties of the system, e.g. the
symmetries. Therefore, it is very important to find particular
solutions for the untruncated system of equations and to adjust them
to track some particular phenomena. In the present paper, we seek
special solutions for the primitive air-motion system. As we should
show later, these solutions can be considered as a model of the
typhoon when it is near its center.\\

Our model is bidimensional and we does not consider the effect of
heating that results from a coupling between the latent heat
released in the clouds and the vertical wind shear. It is known that
a great variety of the vortex phenomenon can be explained in the
frame of two-dimensional models if we consider the vertical
dimension as a small scale comparing with the horizontal ones (e.g.
\cite{DKM}). In meteorology, this concept is often adopted in
studying the processes for middle and large scales \cite{Pedloski}.
In this case, the  atmosphere is regarded to be "thin". For example,
the meteorologists often consider the atmosphere as a one-layer
fluid in the barotropic framework without specifying the vertical
coordinate. This bidimensional theory, despite such simplicity, can
use to explain many observed features of the tropical cyclone
motions \cite{chan}. However, the essence of bidimensionality in our
model differs from above. We derived it from a primitive complete
three-dimensional system of equations by averaging the system over
the height, so the vertical processes of
the system are presented in a hidden form. \\

The paper is organized as follows. In section 2, we study the
derivation of a bidimensional model of the atmosphere dynamics based
on the primitive Navier-Stokes equations for compressible viscous
heat-conductive gas in the physical 3D space. In section 3,  we
investigate polynomial solutions for both barotropic and baroclinic
system and find that the equations which govern the trajectories at
the equilibrium point are the same in both cases. In section 4, we
test possible trajectories of the typhoon center with various input
parameters and demonstrate that our the artificial trajectories can
imitate the behavior of the typhoon eye tracks. Moreover, numerical
simulation shows that the motion of the vortex center provided by
our explicit algebraic formulas is reasonably close to the motion of
a developed vortex when it is near the equilibrium. In section 5, we
examine two different types of historical typhoons: Man Yi
(parabolic trajectory) and Parma (loop), with our theoretical
results. Here, we propose a method of the parameter-fitting that
allows us to predict the trajectory of typhoon eye only based on
three successive points in the past. In section 6, we further
investigate the situation when the typhoon encounters the shore. In
this case, the typhoon may disappear due to the destruction of the
stable vortex. Finally, discussions about our topics and future work
are provided in section 7.

\section{Bidimensional models of the atmosphere dynamics}
The motion of compressible rotating, viscous, heat-conductive, 
Newtonian polytropic gas in ${\mathbb R}\times{\mathbb R}^3,$ is
governed by the compressible Navier-Stokes equations \cite{Landau}
$$\partial_t \rho+{\rm div} (\rho u)=0,\eqno(2.1)$$
$$\partial_t(\rho u)+{\rm Div}(\rho u \otimes u)+\nabla p+2\omega\times u={\rm Div} T,\eqno(2.2)$$
$$\partial_t\left(\frac{1}{2}\rho |u|^2+\rho e\right)+{\rm div}\left((\frac{1}{2}\rho |u|^2+\rho e+p)u
\right)={\rm div}(Tu)+\nabla(\kappa,\nabla \theta), \eqno(2.3)
$$ where $\rho, \, u=(u_1,u_2,u_3),\, p, \, e,\,\theta
\,$ denote the density, velocity, pressure, internal energy and
absolute temperature, respectively. Here
$\omega=(\omega_1,\omega_2,\omega_3)$ is the angular velocity of the
Earth rotation, $\kappa=(\kappa_1,\kappa_2,\kappa_3)^T,\,\kappa_i\ge
0,\,i=1,2,3,$ is the vector of heat conduction.  The stress tensor
$\,T\,$ is given by the Newton law
$$T=T_{ij}=\mu \,(\partial_iu_j+\partial_j u_i)+\lambda \,{\rm div} u\, \delta_{ij},\eqno (2.4)$$
where the constants $\mu$ and $\lambda$ are the coefficient of
viscosity, and the second coefficient of viscosity. Here ${\rm Div}$
and $\rm div$ stand for the divergency of tensor and vector,
respectively, $\otimes$ denotes the tensor product of vectors.

The state equations are
$$p=R\rho \theta,\quad e=c\theta,\quad
p=A\exp\left(\frac{s}{c}\right)\rho^{\tilde\gamma}.\eqno(2.5)
$$
Here $A>0$ is a constant, $R$ is the universal gas constant,
 $s=\log e - (\gamma-1)\log
\rho$ is the specific entropy, $c=\frac{R}{\tilde\gamma-1},$
$\tilde\gamma>1$ is the specific heat ratio.

The state equations (2.5) imply
$$\,p=(\tilde\gamma-1)\rho e,\eqno(2.6)$$
which allows us to consider (2.1) -- (2.3) as a system of equations
for the unknowns $\rho,\,u,\,p.$ Indeed,
 from (2.1) -- (2.3) and (2.6), we arrive
$$\partial_t p +
(u,\nabla p)+\tilde\gamma p\,{\rm
div}u=(\tilde\gamma-1)\,\sum\limits_{i,j=1}^n\,T_{ij}
\partial_j u_i+\frac{\kappa}{R}\,\Delta\frac{p}{\rho}.\eqno(2.7)$$
For simplicity, we denote the system (2.1), (2.2), (2.7) as (NS).

Let ${\bf x}=(x_1,x_2)$ be a point on the Earth surface, $\varphi_0$
be the latitude of some fixed point ${\bf x}_0.$ Following from
\cite{Alishaev} (early this approach was used in \cite{Obukhov} for
barotropic atmosphere), we derive a spatially two-dimensional system
for (NS). Let us introduce $\hat\phi$ and $\bar f$ to represent for
taking the average of $\phi$ and $f$ over the height respectively,
namely, $\displaystyle\hat\phi :=\int_0^\infty \phi\,dz,\quad \bar f
:=\frac{1}{\hat\rho}\int_0^\infty \rho f\,dz.$ where $\phi$ and $f$
are arbitrary functions, and denote $\varrho(t,{\bf x})=\hat\rho,
\,P(t,{\bf x})=\hat p,\,{\bf U}(t,{\bf x})=\bar u,\,{\Theta}(t,{\bf
x})=\bar{\theta}.$ Moreover, the usual adiabatic exponent,
$\tilde\gamma$ is related to the ``two-dimensional" adiabatic
exponent $\gamma$ as follows:
$\gamma=\displaystyle\frac{2\tilde\gamma-1}{\tilde\gamma}<\tilde\gamma.$

Now, we include the impenetrability conditions in our model. These
conditions ensure that the derivatives of the velocity equal to zero
on the Earth surface and a sufficiently rapid decay for all
thermodynamic quantities as the vertical coordinate $z$ approaches
to infinity. In other words, the impenetrability conditions make
sure the boundedness of the mass, energy, and momentum in the air
column. They also provide the necessary conditions for the
convergence of integrals .

Let us denote additionally $ l =2 \omega_3 \sin \varphi_0 $ (the
Coriolis parameter) and $\quad L = \left(\begin{array}{cr} 0 & -1 \\
1 & 0
\end{array}\right).$

In this way, one can get the system as the $l$- plane approximation
near ${\bf x}_0$:
$$
\varrho(\partial_t {\bf U} +  ({\bf U} \cdot \nabla ){\bf U} + l L
{\bf U} + \nabla P) = T_1(\bf U),
$$
$$
\partial_t \varrho +  \nabla \cdot ( \varrho {\bf U}) =0,\eqno(2.8)
$$
$$\partial_t P +
({\bf U},\nabla P)+\gamma P\,{\rm div}{\bf U}=(\gamma-1)\,T_2({\bf
U}) +\kappa (\frac{1}{R}\,\Delta\Theta - \xi),
$$
where
$$T_1({\bf U})= {\rm Div} {\bf T}_{ij},\quad {\bf T}_{ij}=\mu \,(\partial_i{\bf U}_j+\partial_j {\bf U}_i)+\lambda \,{\rm
div} {\bf U}\, \delta_{ij},
$$
$$
T_2({\bf U})=((2\mu+\lambda)((\partial_1{\bf U}_1)^2+(\partial_2{\bf
U}_2)^2)+\mu (\partial_1{\bf U}_2+\partial_2{\bf U}_1)^2),
$$
$\mu$ and $ \lambda $ are  the coefficients of viscosity interpreted
as the coefficients of turbulent viscosity (may be not only
constants). Their values are much greater than those for the
molecular analogues.  For simplicity, the coefficients of the heat
conductivity are assumed to be constants and equal. $\xi$ is the
heat flow from the ocean surface, it is also assumed to be a
constant.

\section{Models of the typhoon dynamics with the linear velocity profile}

\begin{figure}
\centerline{\includegraphics[width=0.6\columnwidth]{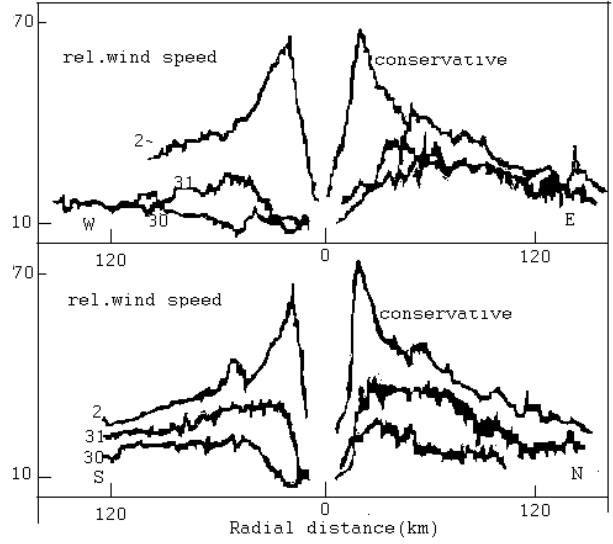}}
\caption{The tangential component of (relative) velocity in $m \cdot
s^{-1}$ on different stage of the typhoon formation (August 30,
August 31, September 2). The hurricane Anita, North Atlantic, 1977
\cite{IAV}}\label{fig1}
\end{figure}

Suppose that the velocity vector near the origin has the form
\begin{equation}\label{3.1}{\bf u}(t,{\bf x})=a(t){\bf r}+b(t){\bf r}_\bot,\end{equation}
where
$${\bf r}=(x_1,x_2)^T,\, {\bf r}_\bot=(x_2,-x_1)^T.$$
This  assumption is made according to the observational data
\cite{IAV}(see Fig.\ref{fig1}). Moreover, as it is shown in
\cite{DST}, in the stable vortex, the velocity field necessarily
satisfies the Cauchy-Riemann conditions, namely,
$$
\frac{\partial u_1}{\partial x_1}=\frac{\partial u_2}{\partial
x_2},\qquad \frac{\partial u_1}{\partial x_2}=-\frac{\partial
u_2}{\partial x_1}.
$$
This property holds for (\ref{3.1}) and also verifies our choice for
the velocity field.

Note that if we do not take into account the turbulent viscosity
($\mu=\lambda=0$) and the heat transfer ($\kappa=0$) in our model.
We get the system similar to the gas dynamics instead of (2.1)-
(2.3). Solutions to such system were constructed and also studied in
\cite{RozSp},\cite{RozPas},\cite{RozSing}.

We next consider two forms of the system $(2.8),$ which have exact
polynomial solutions.

\subsection{System with respect to the velocity, pressure and entropy, without viscosity and heat conductivity}

We assume $\mu=\lambda=\kappa=0$ and introduce a new variable
$\pi=p^{\frac{\gamma-1}{\gamma}}.$
We also consider the 2D entropy $S$ connecting with $\varrho$ and
$P$ through the state equation (2.5), where we use $\gamma $ instead
of $\tilde\gamma.$ For the new unknown variables $\pi, {\bf U}, S,$
we arrive the system \cite{Rrcd}
$$
\partial_t {\bf U} +  ({\bf U} \cdot \nabla ){\bf U} +
l L {\bf U} + \frac{\gamma}{\gamma-1}\exp{\frac{S}{\gamma}}\nabla
\pi =\varrho^{-1} T_1 ( {\bf U}),
$$
$$
\partial_t \pi +  (\nabla \pi \cdot  {\bf U}) +
(\gamma-1) \pi \, {\rm div}  {\bf U}=0, $$
$$\frac{p}{R}\left(\partial_t S +({\bf U},\nabla
S)\right)=T_2({\bf U}).$$

 Following \cite{BVDD,BVDD1}, we change the
coordinate system so that the origin of the new system is located at
the center of the typhoon eye. Now ${\bf U}= {\bf u}+{\bf V},$ where
${\bf V}(t)=(V_1(t),V_2(t))$ is the velocity of the typhoon eye
propagation. Thus, we obtain a new system
\begin{equation}\begin{aligned} \label{2.1}
\partial_t {\bf u} + ({\bf u} \cdot \nabla ){\bf u} +
\dot {\bf V}+ l L ({\bf u}+{\bf V})
+\frac{\gamma}{\gamma-1}\exp{\frac{S}{\gamma}}\nabla \pi =
\varrho^{-1} T_2({\bf u}),\end{aligned}
\end{equation}
\begin{equation}\label{2.2}
\partial_t \pi +  \nabla \pi\cdot {\bf u} + (\gamma-1)\pi{\rm div}{\bf u}
= 0,  \end{equation}
\begin{equation}\label{2.3}
\frac{p}{R}\left(\partial_t S +({\bf u},\nabla S)\right)=T_2({\bf
u}).
\end{equation}

Note that in \cite{BVDD,BVDD1}, shallow water equations were
considered. It is a particular case of (2.8) with
$\gamma=2,\,S\equiv const., \lambda=\mu=\kappa=0.$

Given the vector ${\bf V}$, the trajectory can be found by
integrating the system \begin{equation}\label{3.4}\dot
x_1(t)=V_1(t),\qquad \dot x_2(t)=V_2(t).\end{equation}

We assume the velocity has a linear profile (\ref{3.1}) and look for
other components of the solution to the system
(\ref{2.1})--(\ref{2.3}) in the following form
\begin{equation}
\label{3.11} \pi(t,{\bf x})=A(t)x_1^2+B(t)x_1x_2+C(t)x_2^2+M(t)x_1+
N(t)x_2+K(t),
\end{equation}
\begin{equation}
\label{3.21} S(t,{\bf x})=S_0(t)+\gamma\,\ln
\sum\limits_{i_1,i_2=1}^{\infty}\,S_{i_1 i_2}(t)x_1^{i_1} x_2^{i_2}.
\end{equation}
Here, $T_1( {\bf u})=T_2( {\bf u})=0.$  Recall that $K(t)$ is the
value of ``renormalized" pressure $\pi$ at the center of vortex,
therefore, in the physical sense, $K(t)>0.$  Substituting
(\ref{3.1})--(\ref{3.21}) into (\ref{2.1})--(\ref{2.3}) and match
the coefficients for similar term, we obtain $S_{i_1 i_2}(t)\equiv
0, $ $\,i_1, i_2\in \mathbb N,\,S_0(t)=const,\, A(t)=C(t),$ and
$\,B(t)\equiv 0.$ At the center of typhoon, there exists a domain of
lower pressure. Therefore, it is natural to set $A(t)>0.$

Note that the motion near the typhoon center is ``barotropic" in our
solution, i.e., the pressure only depends on its density. However,
this barotropicity is only for the bidimensional density and
pressure. This assumption may not hold for three-dimensional cases.

Let us introduce a constant $
c_0:=\frac{\gamma}{\gamma-1}\exp{\frac{S_0}{\gamma}}.$

Thus, the functions $a(t), b(t), A(t), M(t), N(t), K(t), V_1(t),
V_2(t)$ satisfy the following system of ODEs:
\begin{equation}\label{3.41}\dot A+2\gamma
aA=0,\end{equation}\begin{equation} \dot a+a^2-b^2+lb+2c_0
A=0,\label{3.51}\end{equation}
\begin{equation}\dot b+2ab-la=0,\label{3.61}\end{equation} \begin{equation}\dot
K+2(\gamma-1)aK=0,\label{3.7}\end{equation}
\begin{equation}\dot M+(2\gamma-1)aM-bN=0,\label{3.8}\end{equation}
\begin{equation}\dot N+(2\gamma-1)aN+bM=0,\label{3.9}\end{equation}
\begin{equation}\dot V_1-lV_2+c_0 M=0,\label{3.10}\end{equation}
\begin{equation}\dot V_2+lV_1+c_0 N=0.\label{3.11}\end{equation}

 From (\ref{3.41}) and (\ref{3.61}), we have
\begin{equation}b=\frac{l}{2}+C_1|A|^{1/\gamma},\label{3.12}\end{equation} where
$C_1.$ a constant. Therefore, system (\ref{3.41}) -- (\ref{3.61})
can be reduced to equations \begin{equation}
\label{3.13}\dot A=-2\gamma aA,\qquad \dot
a=-a^2-\frac{l^2}{4}+C_1^2A^{2/\gamma}-2c_0 A.
\end{equation}

Further, if $A(t)$ and $a(t)$ are known, we can find other
components of solutions. Namely, from (\ref{3.8}) and (\ref{3.9}), c
we can get
$$M(t)=(M^2(0)+N^2(0))^{1/2}\exp\left(-\frac{2\gamma-1}{2}\int\limits
_0^t a(\tau) d\tau\right) \sin(\frac{l}{2}t+C_1\int_0^t
A^{1/\gamma}(\tau)d\tau +C_2),$$
$$N(t)=(M^2(0)+N^2(0))^{1/2}\exp\left(-\frac{2\gamma-1}{2}\int\limits_0^t
a(\tau) d\tau\right) \cos(\frac{l}{2}t+C_1\int_0^t
A^{1/\gamma}(\tau)d\tau +C_2).$$  From (\ref{3.41}) and (\ref{3.7}),
we obtain
$$K(t)=C_3(|A(t)|)^{\frac{\gamma-1}{\gamma}}.$$  Here $C_2, C_3$ are
constants depending only on initial data. However, from
(\ref{3.10}), (\ref{3.11}), and (\ref{3.4}), the trajectory does not
depend on $K(t).$


The phase curves of (\ref{3.13}) can be found explicitly. They
satisfy the algebraic equation \begin{equation}a^2=C_4
A^{\frac{1}{\gamma}}-C_1^2A^{\frac{2}{\gamma}}+\frac{l^2}{4(\gamma-1)}+
\frac{2c_0}{\gamma-1}A,\label{3.14}\end{equation} with the constant
$C_4$ depending only on initial data.  Our analysis shows that there
exists a unique equilibrium point on the phase plane $(A,a),\, A>0$
for $1<\gamma<2$. It is a center on the axis $a = 0.$ We denote this
stable equilibrium as $(A_0,0),$ where $A_0$ is a positive root of
the equation $-\frac{l^2}{4}+C_1^2A^{2/\gamma}-2c_0 A=0.$


\subsection{System with respect to the velocity, density  and temperature}

We take the average of the first state equations (2.5) over the
height to get $P=R\varrho \Theta$ and then exclude the pressure from
system (2.8), we arrive
$$ \varrho(\partial_t {\bf U} +  ({\bf U} \cdot
\nabla ){\bf U} + l L {\bf U}) + R(\varrho\nabla \theta + \theta
\nabla \varrho) = T_1( {\bf U}),
$$
$$
\partial_t \varrho +  (\nabla \varrho \cdot  {\bf U}) +
\varrho \nabla  {\bf U}=0, $$
$$
\partial_t \Theta +  (\nabla \Theta \cdot  {\bf U}) +
(\gamma-1)\,\Theta \,{\rm div }{\bf U}=\frac{\gamma-1}{R} T_2({\bf
U})+\frac{\kappa}{R \varrho}(\Delta \theta-\xi). $$ After changing
the coordinates system, it takes the form
\begin{equation}\begin{aligned} \label{3.1a}
\varrho(\partial_t {\bf u} + ({\bf u} \cdot \nabla ){\bf u} + \dot
{\bf V}+ l T ({\bf u}+{\bf V}))+ R(\varrho\nabla \theta + \theta
\nabla \varrho)= T_1( {\bf u}),\end{aligned}
\end{equation}
\begin{equation}\label{3.2a}
\partial_t \varrho +  (\nabla \varrho \cdot  {\bf u}) +
\varrho\,{\rm div}  {\bf u}=0 ,  \end{equation}
\begin{equation}\label{3.3a}
\partial_t \Theta + ( \nabla \Theta \cdot  {\bf u}) +
(\gamma-1)\,\Theta \,{\rm div}  {\bf u}\,=\,\frac{\gamma-1}{R}
T_2({\bf u})+\frac{\kappa}{R\varrho}(\Delta \theta-\xi).
\end{equation}
The position of the vortex center can be obtained again from
(\ref{3.4}).

Note that now the viscosity and heat conductivity are not neglected.

We can obtain a closed system of ordinary differential equations if
we seek the solution to (\ref{3.1a})--(\ref{3.3a}) in the following
form:
\begin{equation}
\label{3.21a} \theta(t,{\bf
x})=A_1(t)x_1^2+B_1(t)x_1x_2+C_1(t)x_2^2+M_1(t)x_1+
N_1(t)x_2+K_1(t),
\end{equation}
\begin{equation}\varrho(t,{\bf
x})=K_2(t) \label{3.31a} .
\end{equation}
Substituting (\ref{3.1}), (\ref{3.21a}), (\ref{3.31a}) into
(\ref{3.1a})--(\ref{3.3a}), we obtain $A_1(t)=C_1(t),$ $B_1(t)=0$
and
\begin{equation}\label{3.4b}\dot A_1+2\gamma
aA_1=0,\end{equation}
\begin{equation} \dot a+a^2-b^2+lb+4R
A_1=0,\label{3.5b}\end{equation}
\begin{equation}\dot b+2ab-la=0,\label{3.6b}\end{equation}
\begin{equation}\dot
K_2+2aK_2=0,\label{3.7a}\end{equation}
\begin{equation}\dot
K_1+2(\gamma-1)aK_1\,=\,\frac{2(\gamma-1)(2\mu+\lambda)}{R}\,a^2+\frac{\kappa}{K_2}(\xi-4A_1),\label{3.7b}\end{equation}
\begin{equation}\dot M_1+(2\gamma-1)aM_1-bN_1=0,\label{3.8a}\end{equation}
\begin{equation}\dot N_1+(2\gamma-1) aN_1+bM_1=0,\label{3.9a}\end{equation}
\begin{equation}\dot V_1-lV_2+2 R M_1=0,\label{3.10a}\end{equation}
\begin{equation}\dot V_2+lV_1+2R N_1=0.\label{3.11a}\end{equation}

Equations (\ref{3.4b}), (\ref{3.5b}), (\ref{3.6b}) can be reduced to
the system of equations as before:
\begin{equation}
\label{3.13a}\dot A_1=-2\gamma aA_1,\qquad \dot
a=-a^2-\frac{l^2}{4}+\bar C_1^2A_1^{2/\gamma}-4 R A_1,
\end{equation}
\begin{equation}b=\frac{l}{2}+\bar C_1|A_1|^{1/\gamma},\label{3.12a}\end{equation}
where $\bar C_1 $ is a constant.

Assuming that we know $a(t)$ and $A_1(t),$ we can find $b(t),\,
M_1(t),\,N_1(t),\,V_1(t)$ and $V_2(t)$ as before. The only stable
equilibrium on the phase plane $(a, A_1) $ for $\gamma\in (1,2)$ is
a center $(0, \bar A_0),$ where $\bar A_0$ is a root of the equation
$\,-\frac{l^2}{4}+\bar C_1^2A_2^{2/\gamma}-4 R A_2=0.$

It is important that the viscosity, the heat conduction and heat
inflow terms show only in the equation (\ref{3.7b}), therefore,
their values only affect the function $K_1(t)$. Nevertheless, the
shape of trajectory defined by the equations (\ref{3.4b}) -
(\ref{3.6b}), (\ref{3.8a}) - (\ref{3.11a}) does not depend on
$K_1(t).$

\section{Possible trajectories}

Let us analyze trajectories of a stable typhoon for both
``barotropic" and ``baroclinic" cases. At the equilibrium point,
$a(t)=0.$ Other coordinates of the stable point for the
``barotropic" system (\ref{3.41}),(\ref{3.51}), (\ref{3.61}) are
$A(t)=A_0$ and $b(t)=b_0= \displaystyle \frac{l}{2}+C_1(
A_0)^{\frac{1}{\gamma}}.$ In the real meteorological situations,
$l>> b_0.$ Thus, from (\ref{3.8} -- \ref{3.11}) and (\ref{3.4}), we
obtain
\begin{align*}x_1(t)= & \quad x_1(0)+\frac{V_2(0)}{l}+\frac{c_0 M(0)}{b_0
l}&{\rm(4.1)}&
\\+&\quad \left(\frac{V_1(0)}{l}-\frac{c_0 N(0)}{l(b_0-l)}\right)\sin lt
-\left(\frac{V_2(0)}{l}+\frac{c_0 M(0)}{l(b_0-l)}\right)\cos lt&&\\+
& \quad \frac{c_0 N(0)}{b_0(b_0-l)}\sin b_0t+\frac{c_0
M(0)}{b_0(b_0-l)}\cos b_0t,&&\\x_2(t)= & \quad
x_1(0)-\frac{V_1(0)}{l}+\frac{c_0 N(0)}{b_0 l}&{\rm(4.2)}&
\\+ &\quad\left(\frac{V_2(0)}{l}+\frac{c_0 M(0)}{l(b_0-l)}\right)\sin lt
+\left(\frac{V_1(0)}{l}-\frac{c_0 N(0)}{l(b_0-l)}\right)\cos
lt&&\\&\quad -\frac{c_0 M(0)}{b_0(b_0-l)}\sin
b_0t+\frac{c_0N(0)}{b_0(b_0-l)}\cos b_0t,&&\end{align*}

for $l\ne b_0.$ The analogous results we get in the ``baroclinic"
case at the stable equilibrium for the system (\ref{3.4b}),
(\ref{3.5b}), and (\ref{3.6b}), namely, $a(t)=0,\,A_1(t)=\bar A_0,
b(t)=b_0=\displaystyle \frac{l}{2}+\bar C_1( \bar
A_0)^{\frac{1}{\gamma}}$. To obtain the equations for the
trajectories, we should change $c_0 M(0)$ to $2RM_2(0)$ and $c_0
N(0)$ to $2R N_2(0).$

\begin{figure}
\centerline{\includegraphics[width=0.6\columnwidth]{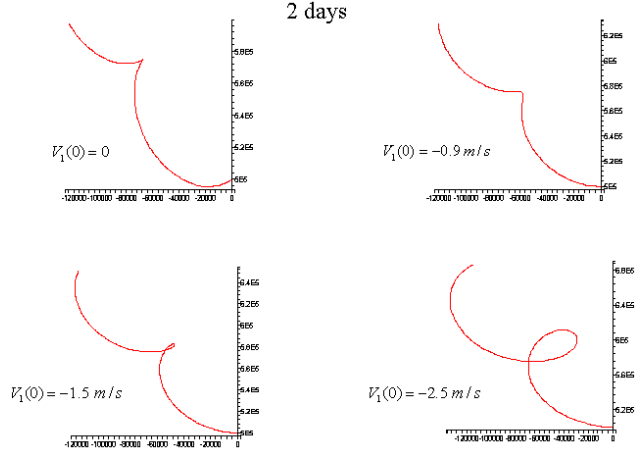}}
\caption{Possible artificial trajectories at $22^o
N,\,$$b_0=10^{-6}\,s^{-1},\,$
$c_0=10^{-1},\,$$M(0)=N(0)=10^{-9},\,$$V_2(0)=0$\,(in corresponding
units) \cite{Rrcd} }\label{fig2}
\end{figure}

\begin{figure}
\centerline{\includegraphics[width=0.6\columnwidth]{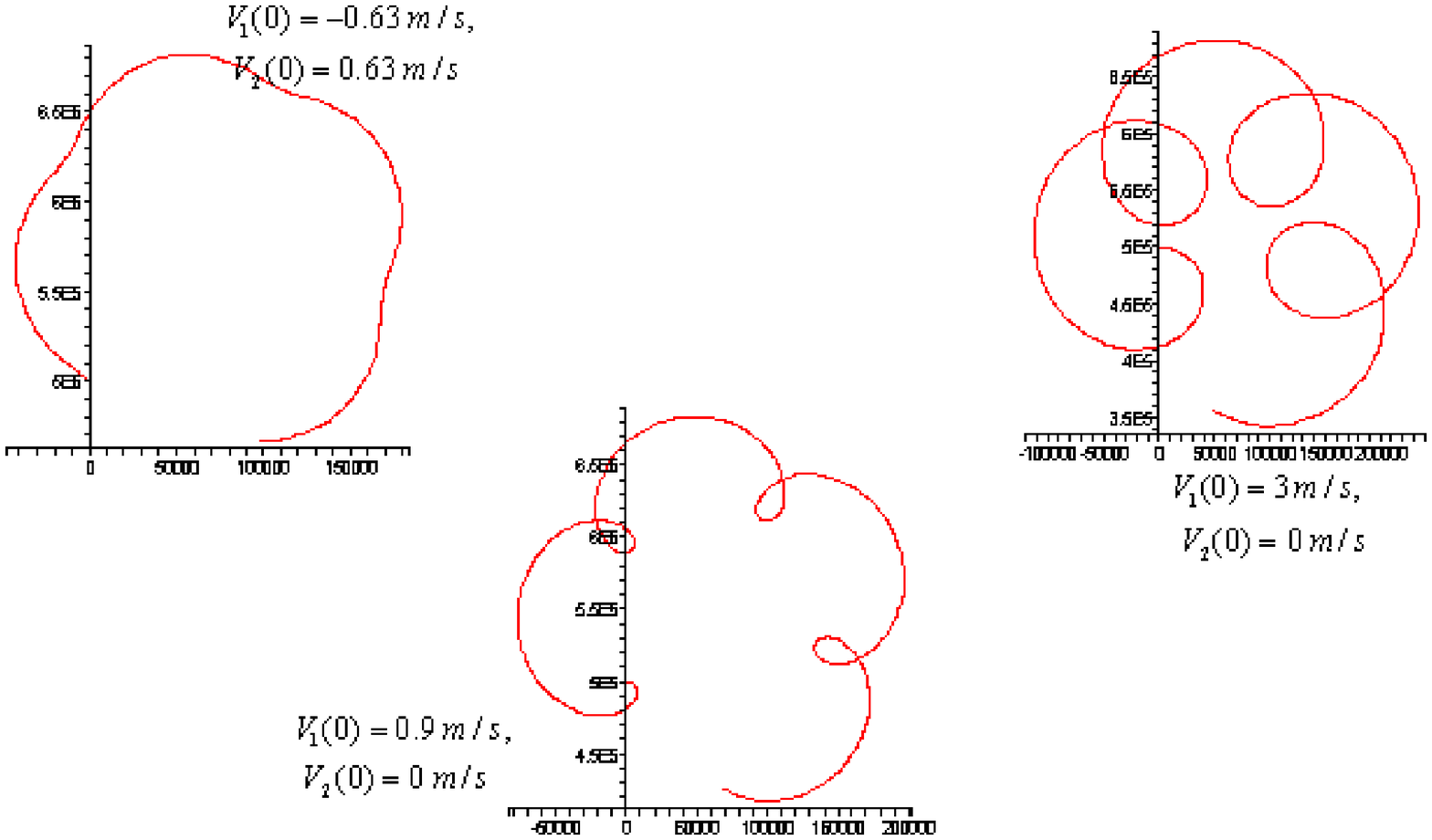}}
\caption{Possible artificial trajectories at $22^o
N,\,b_0=10^{-5}\,s^{-1},\,$ $c_0=10^{-1},\,$$M(0)=N(0)=10^{-9}$\,(in
corresponding units) \cite{Rrcd}}\label{fig3}
\end{figure}

Thus, the trajectory is a superposition of two circular motions: one
of them has a period $2\pi/l,$ the other one has a period
$2\pi/b_0.$ They can lead to the appearance of loops, sudden change
of directions and other complicated trajectories. Several examples
of artificial trajectories are presented in Figs.\ref{fig2} and \ref{fig3}.

\medskip

\begin{figure}
\centerline{\includegraphics[width=0.6\columnwidth]{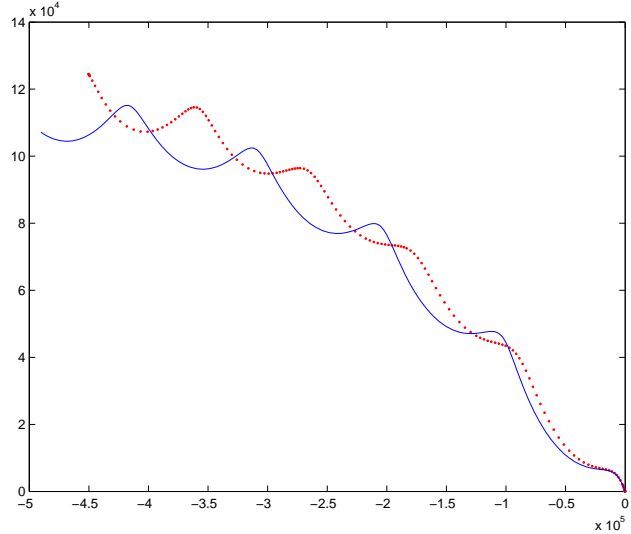}}
\caption{Difference between trajectories corresponding to the exact
equilibrium point (solid line) and in a vicinity of equilibrium
(dashed line) within three days }\label{fig4}
\end{figure}

It is possible to compute the position $(x_1, x_2)$ of the
trajectory based on the system (\ref{3.41} -- \ref{3.11}),
(\ref{3.4}) by applying ODE integrators. However, the difference
between the numerical trajectory and points on the curve (4.1),
(4.2) for the physical values of parameters is very little. For
example, for the initial data $a(0)=10^{-5} s^{-1},\, b(0)=5\cdot
10^{-5}s^{-1},\, A(0)=10^{-9}\,(N\cdot
m^{-2})^{(\gamma-1)/\gamma}\cdot m^{-2},\, N(0)=10^{-3}\,(N\cdot
m^{-2})^{(\gamma-1)/\gamma}\cdot
m^{-1},\,M(0)=2\,N(0),\,V_1(0)=-1\cdot m\cdot s^{-1},\,
V_2(0)=1\cdot m\cdot s^{-1},\,x_1(0)=x_2(0)=0,$ the difference in
positions after three days is about $20\,km $ (see Fig.\ref{fig4}).
(Here, the radius of typhoon is taken as 300 $km,$ the difference
between the pressure at the center of typhoon and the ambient
pressure is $10^{4}\,Pa,$ the constant $c_0$ is estimated as
$10^{-1}$).  Therefore, basically, there is no reason to get the
solution curves by applying numerical integration for (\ref{3.41} --
\ref{3.11}) and (\ref{3.4}) instead of using explicit formulas (4.1)
and (4.2).

\section{Examples of analysis of  real trajectories}

Assume initial conditions are known, one can integrate the system
(\ref{3.41})-- (\ref{3.11}), (\ref{3.4}). But these initial
conditions include the {\it height-averaged} physical quantities
such as divergency $a(0),$ vorticity $b(0),$ parameters $A(0),$ $
M(0),$ $ N(0),$ (relating to the pressure field), and the initial
velocity $V(0).$ It is not clear how to measure these quantities.
The only widely available information is positions of the typhoon
eye. Thus, we try to fit the parameters using the formulas (4.1) and
(4.2). However, it needs to assume that the vortex is at its
conservative phase.

We use the following algorithm: at the first step, we choose three
successive points of trajectory   ($(x_1(t_0),x_2(t_0)),
(x_1(t_1),x_2(t_1)),(x_1(t_2),x_2(t_2))$). Formulas (4.1), (4.2)
give a system of four linear algebraic equations with respect to
$V_1(0),\,V_2(0),\,c_0M(0),\,c_0 N(0).$ The parameter $b_0$ is free.
The value of the Coriolis parameter $l$ is calculated at the initial
point $(x_1(t_0),x_2(t_0)).$ The second step is to choose the
parameter $b_0.$ If we deal with historical trajectory, $b_0$ can be
fitted so that the artificial trajectory coincides with the real one
as long as possible. As we should show, the coincidence sometime
lasts for several days.

In practice, we can compute $b_0,$ for example, from the condition
\begin{equation} \label{b_01} V_1(0)=\bar
V_1(0),\end{equation}
\begin{equation} \label{b_02} V_2(0)=\bar V_2(0),\end{equation} where $\bar
V_i(0)=\frac{x_i(t_1)-x_i(t_0)}{t_1-t_0},\,i=1,2.$ However,
(\ref{b_01}) and (\ref{b_02}) give different values of $b_0$
basically.

Let us estimate all possible values of parameter $b_0$ at the
equilibrium point.  As it follows from (\ref{3.51}), $b_0$ is a root
of the quadratic equation
$$
b_0^2-lb_0-2c_0A_0=0,
$$
therefore, $b_0\approx -\frac{2c_0A_0}{l}.$ for $c_0A_0<<l^2$. If we
further assume $A_0=10^{-9},\, c_0=10^{-1},\,l=10^{-4},$ $b_0\approx
-2\cdot 10^{-6}.$ Thus, to be physically meaningful, we can assume
$|b_0|\le 10^{-5}$ and bear in mind that this value is essentially
negative.

Thus, if the set of three points is not able provide an appropriate
value of $b_0,$ the prediction is considered as a failure. We need
to shift these three points until we get a definite value of $b_0$
that can be used for the weather forecast.

We consider three successive points as suitable candidates for a
forecast if equations (\ref{b_01}) and (\ref{b_02}) have roots
$b_{01}$ and $ b_{02}$ in the interval $|b_0|\le 10^{-5}$
respectively and $|b_{01}-b_{02}|<\varepsilon,$ with $\varepsilon$
sufficiently small. In this case, we chose $b_0$ as the mean of
$b_{01}$ and $b_{02}.$

In the next, we give two examples of analysis of the real tropical
typhoon tracks.

\subsection{Man Yi}

Let us consider a recent typhoon of the West Pacific region, Man Yi
(4 category, July 8-15, 2007). The historic trajectory and
observational data are given on Figs.5 and 6 \cite{data}. Remark
that the numeration cited in Fig.6 includes certain intermediate
points (like 1A, 2A, etc). Here, We did numerate the intermediate
points as in Fig. 6, instead, we adopt successive numeration in our
examples.

\begin{figure}
\centerline{\includegraphics[width=0.5\columnwidth]{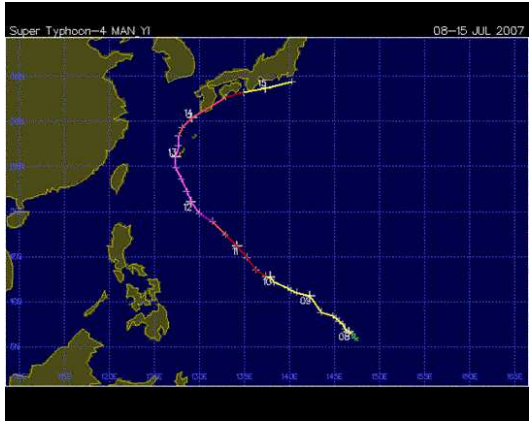}}
\caption{The Man Yi trajectory \cite{data} }\label{fig5}
\end{figure}
\begin{figure}
\centerline{\includegraphics[width=0.8\columnwidth]{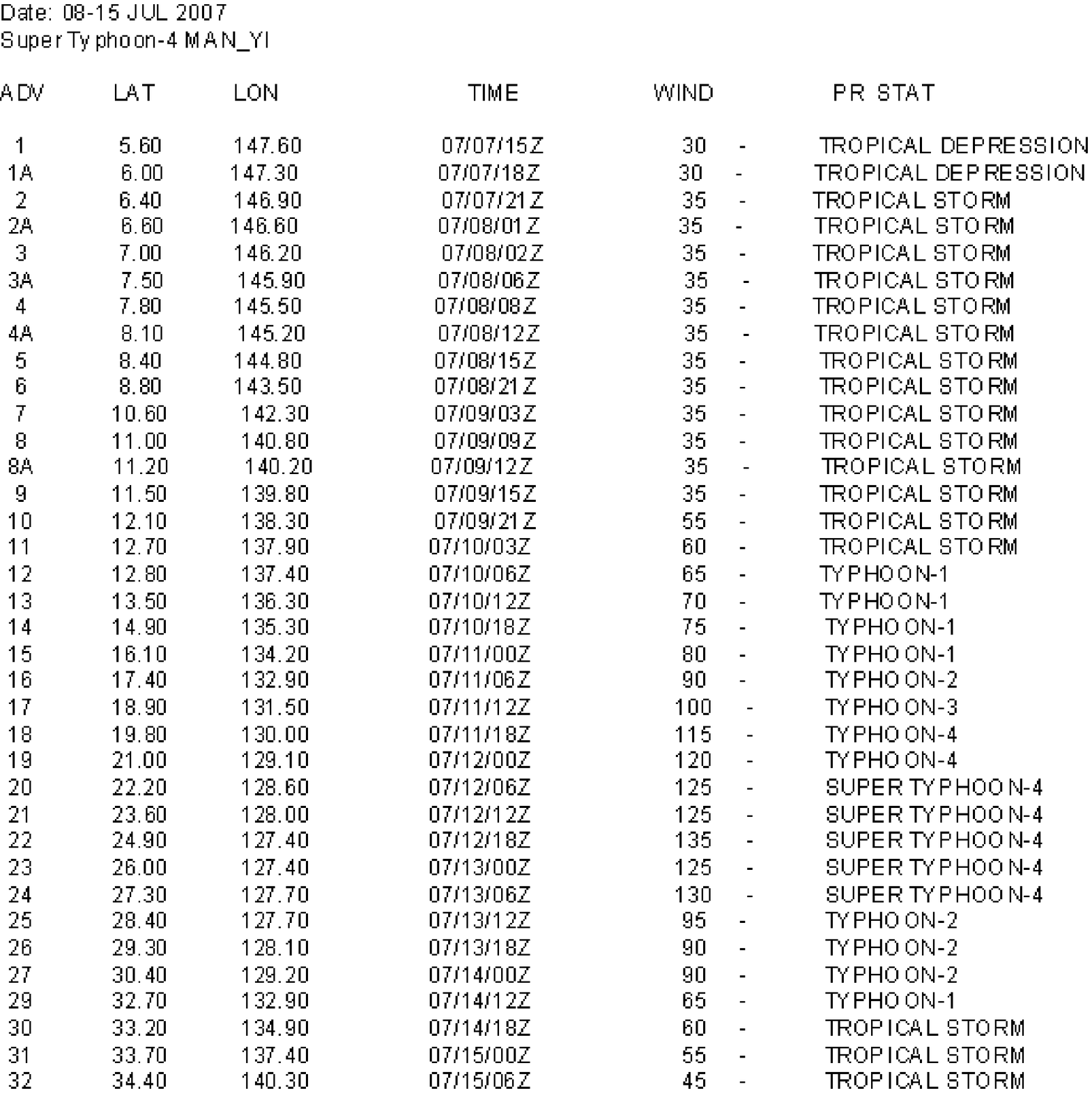}}
\caption{The Man Yi data \cite{data} }\label{fig6}
\end{figure}
\begin{figure}
\centerline{\includegraphics[width=0.5\columnwidth]{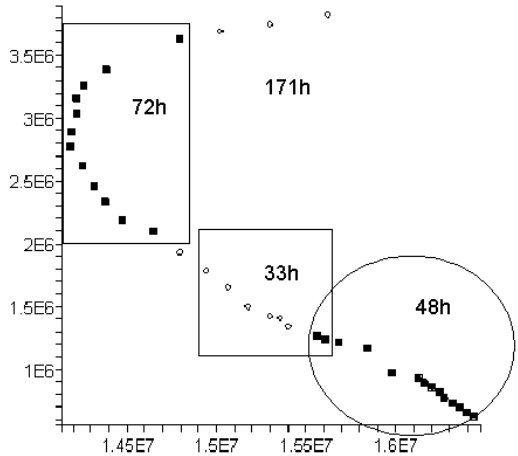}}
\caption{The Man Yi trajectory (observational data) with analyzed
regions}\label{fig7}
\end{figure}

\subsubsection{Historical trajectory: fitting averaged vorticity}

\begin{figure}[h]
\begin{minipage}{0.5\columnwidth}
\centerline{\includegraphics[width=0.7\columnwidth]{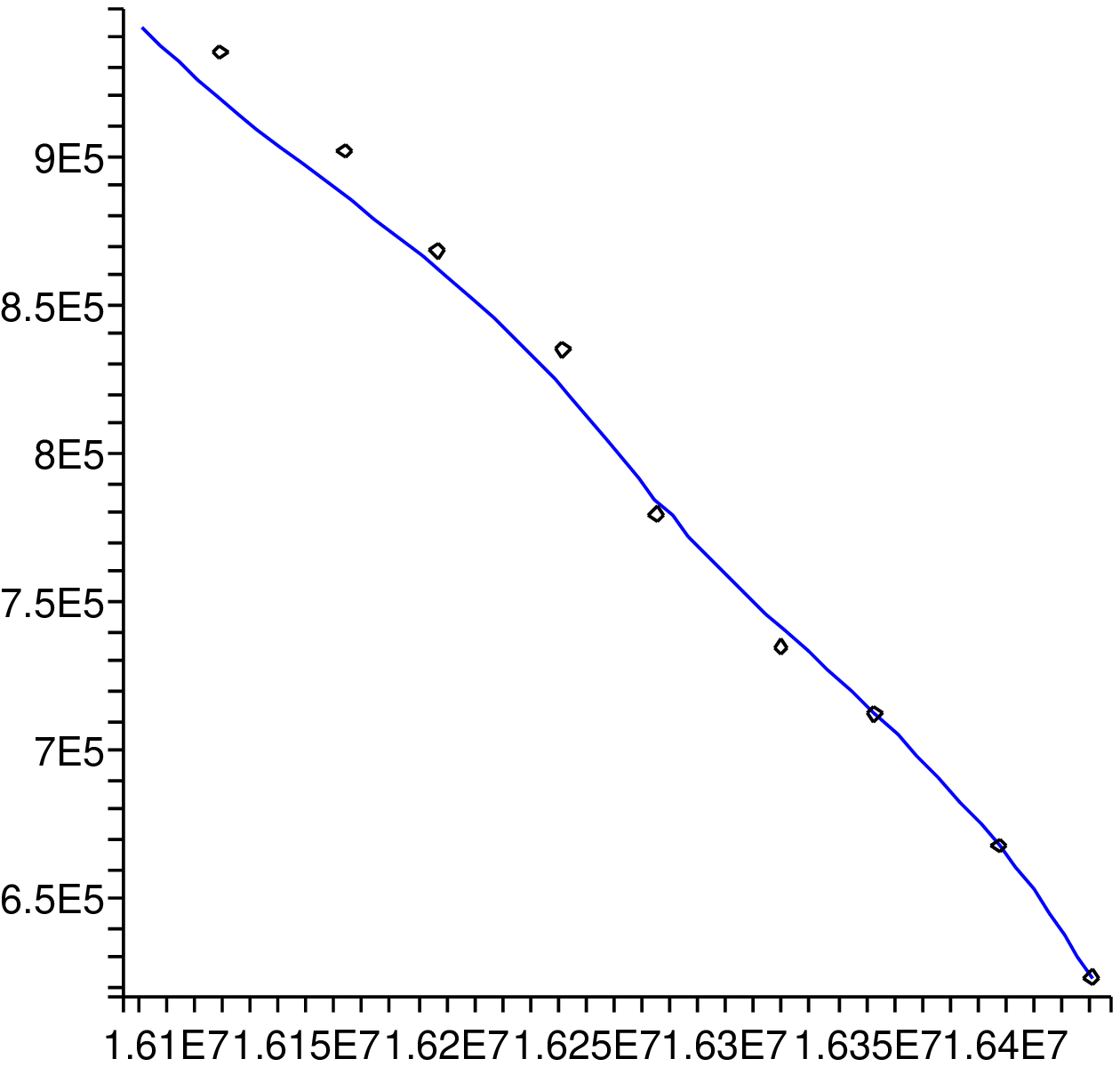}}
\caption{The first 9 points (18 hours). Here and below dots
corresponds to observational data, solid line represents analytic
results }\label{fig8}
\end{minipage}%
\begin{minipage}{0.5\columnwidth}
\centerline{\includegraphics[width=0.7\columnwidth]{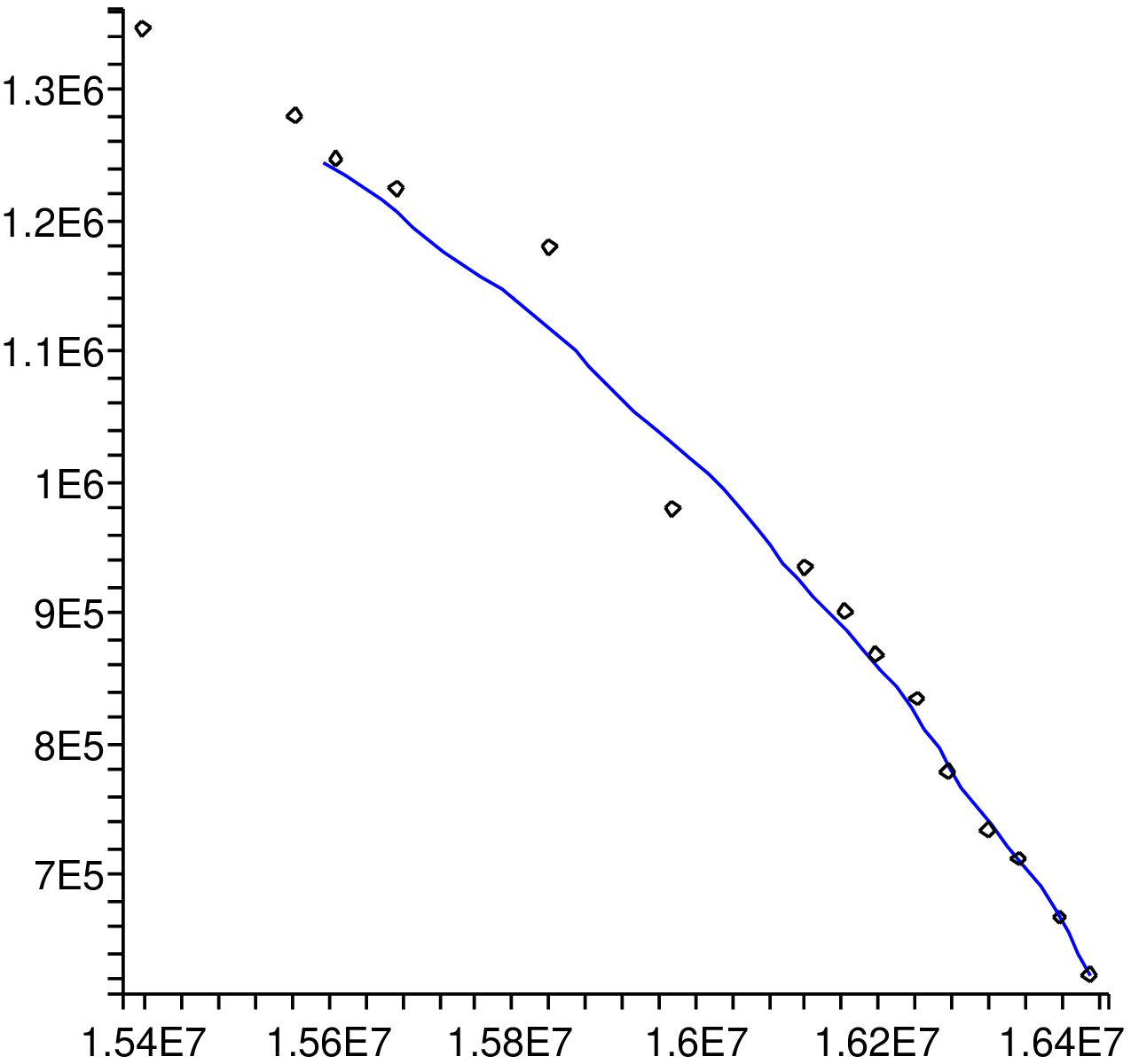}}
\caption{The first 14 points (48 hours) }\label{fig9}
\end{minipage}
\begin{minipage}{0.5\columnwidth}
\centerline{\includegraphics[width=0.7\columnwidth]{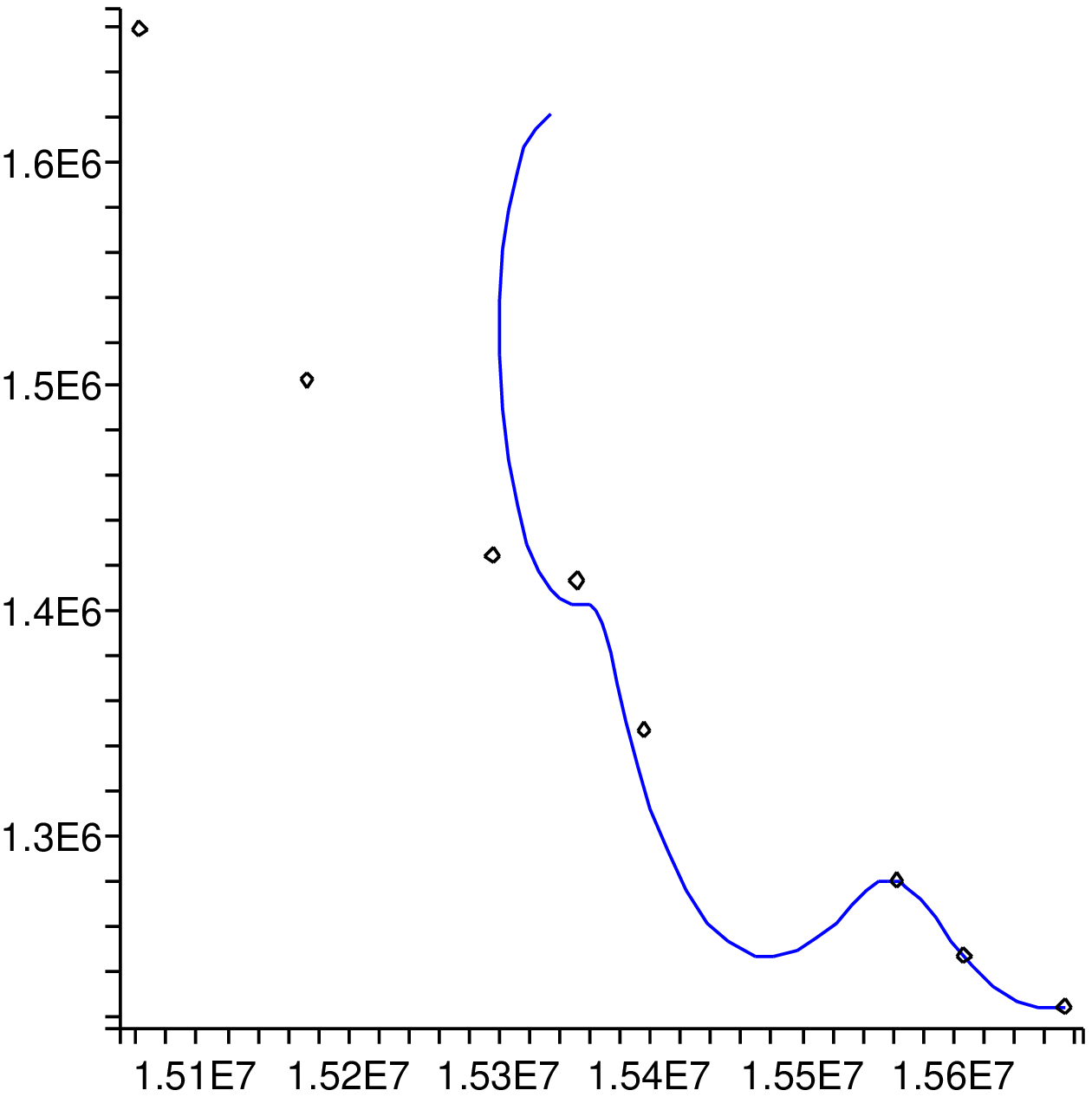}}
\caption{12--19 points (33 hours) }\label{fig10}
\end{minipage}%
\begin{minipage}{0.5\columnwidth}
\centerline{\includegraphics[width=0.7\columnwidth]{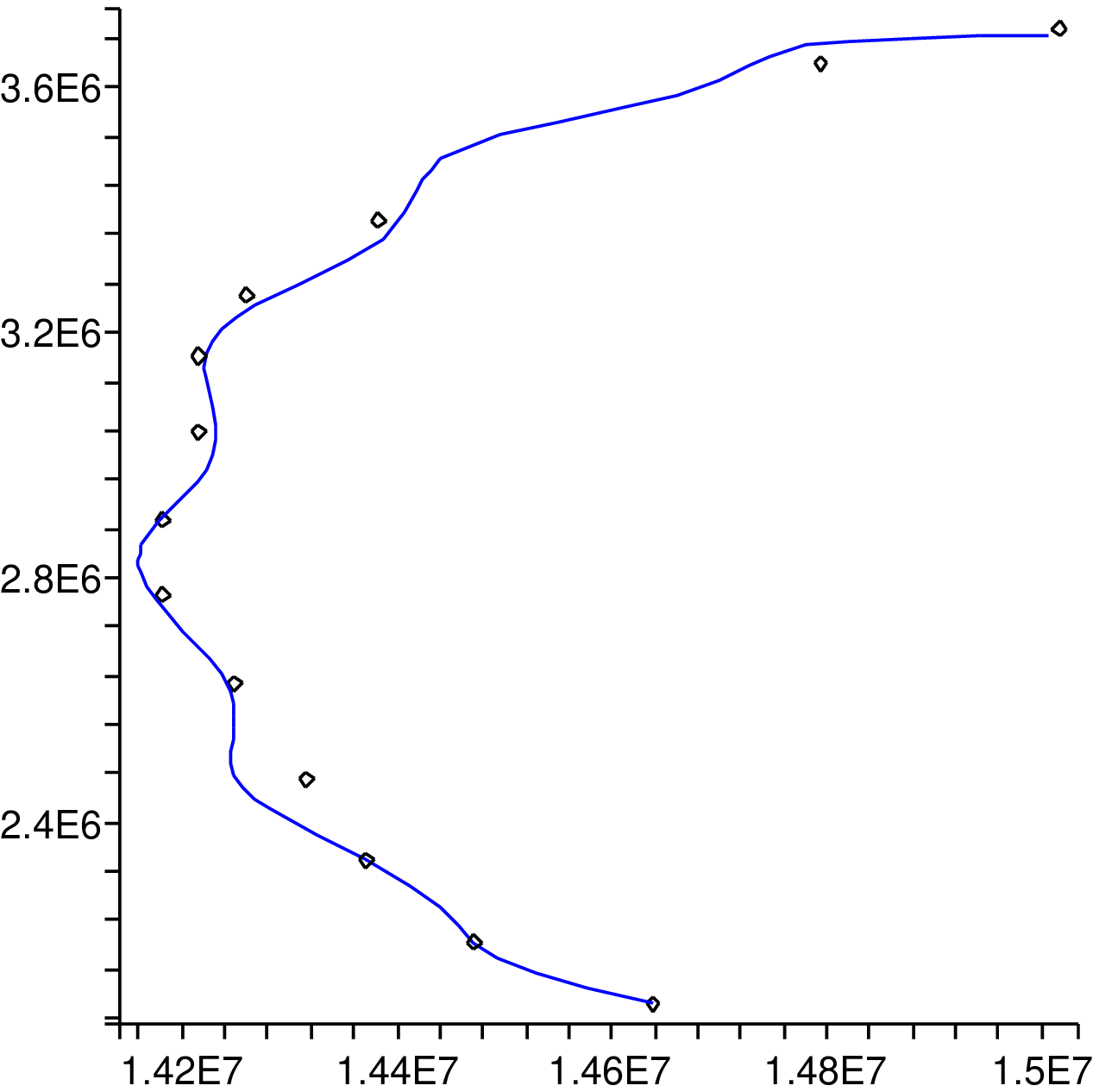}}
\caption{22--34 points (72 hours) }\label{fig11}
\end{minipage}
\end{figure}

By analyzing the trajectory of the typhoon Man Yi, we found that it
is possible to fit $b_0$ so that there is a good coincidence with
the artificial trajectory given by formulas (4.1) and (4.2) in three
different regions (Fig.\ref{fig7}). Here and below graphs, the
diamond shape points represent the observational path of tropical
typhoon and solid lines represent the analytic results; the first
three points of trajectories, where the points and line coincide,
are used for the parameters estimation. The first region corresponds
to the first 14 points on Fig.\ref{fig7} (the total duration is 48
hours, the "tropical storm" stage). Fig.\ref{fig8} shows the
prediction for point 1 to 9 (18 hours), Fig.\ref{fig9} - for point 1
to 14 (48 hours). Here, we get $b_0=2.236198023\cdot
10^{-6}\,s^{-1}.$ The second region on Fig.\ref{fig7} corresponds to
points 12 - 19 (33 hours, "tropical storm - typhoon 1",
Fig.\ref{fig10}). At this region, the vortex is developing, thus we
can not consider it as a stable one. Generally, making predictions
according formulas (4.1), (4.2) may not give good results for
unstable regions. Nevertheless, these formulas with $b_0=-1.8\cdot
10^{-5}\,s^{-1}$ still gives a satisfactory coincidence for the real
and artificial trajectories. The third region that we analyze is
from point 22 to 34, which lasts for 72 hours. In this case, the
typhoon is developing from 3th to 4th stage and then decays. We get
$b_0=-1\cdot 10^{-5}$ and obtain a very good coincidence for the 3
day period.

\subsubsection{Attempt of a forecast in real time}

Now, suppose that that we do not know the track of the whole
trajectory in advance and try to find $b_0$ to compare its value
found from (\ref{b_01}) and (\ref{b_02}). It is worth to note that
one may propose other methods to find the averaged vorticity. This
will be a subject of our further investigations. Let us mention that
one can consider the neural network approaches for this
direction\cite{RT}.

\begin{figure}[h]
\begin{minipage}{0.5\columnwidth}
\centerline{\includegraphics[width=0.7\columnwidth]{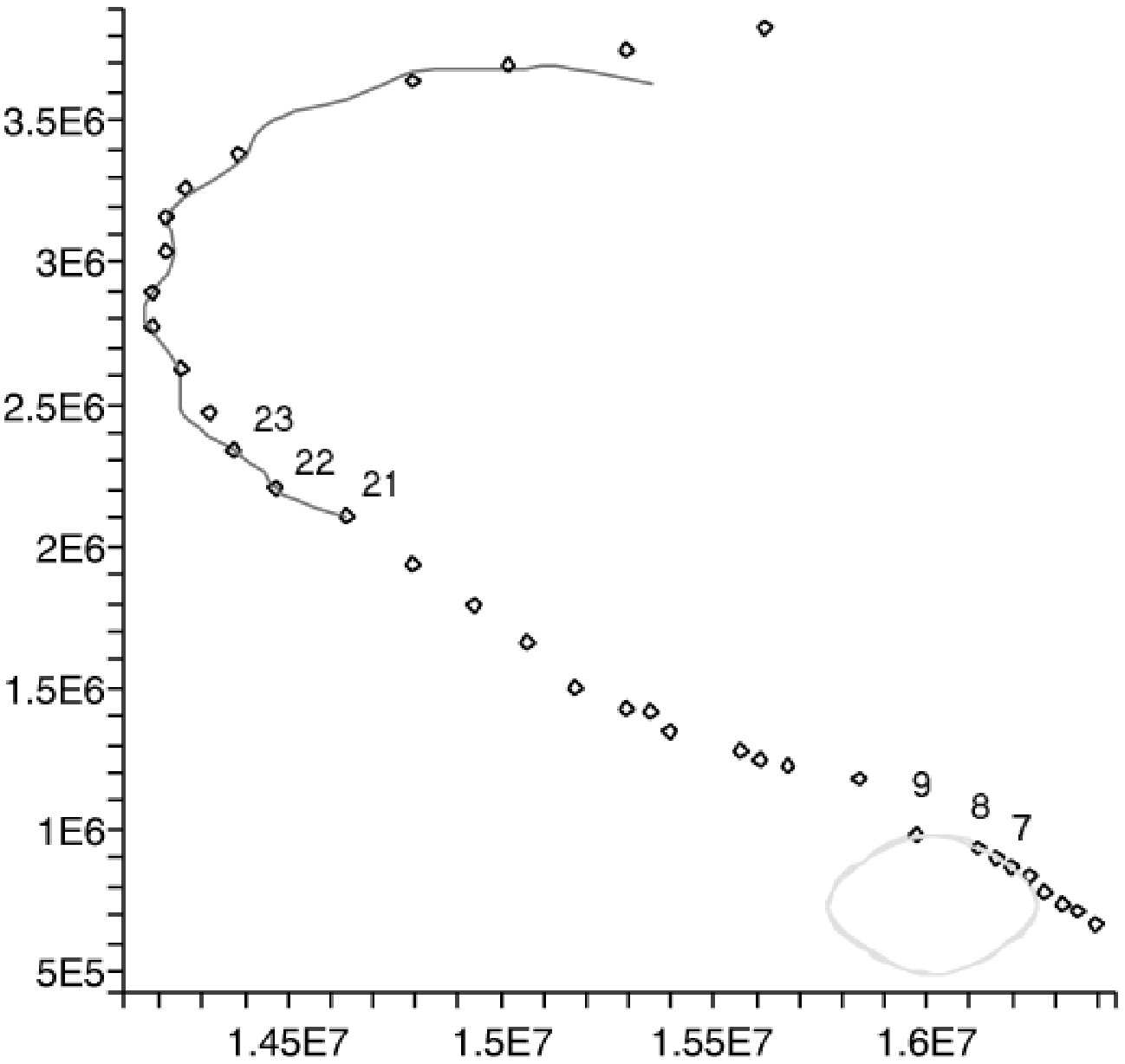}}
\caption{$\varepsilon=2\cdot 10^{-6}$: potential loop }\label{fig12}
\end{minipage}%
\begin{minipage}{0.5\columnwidth}
\centerline{\includegraphics[width=0.7\columnwidth]{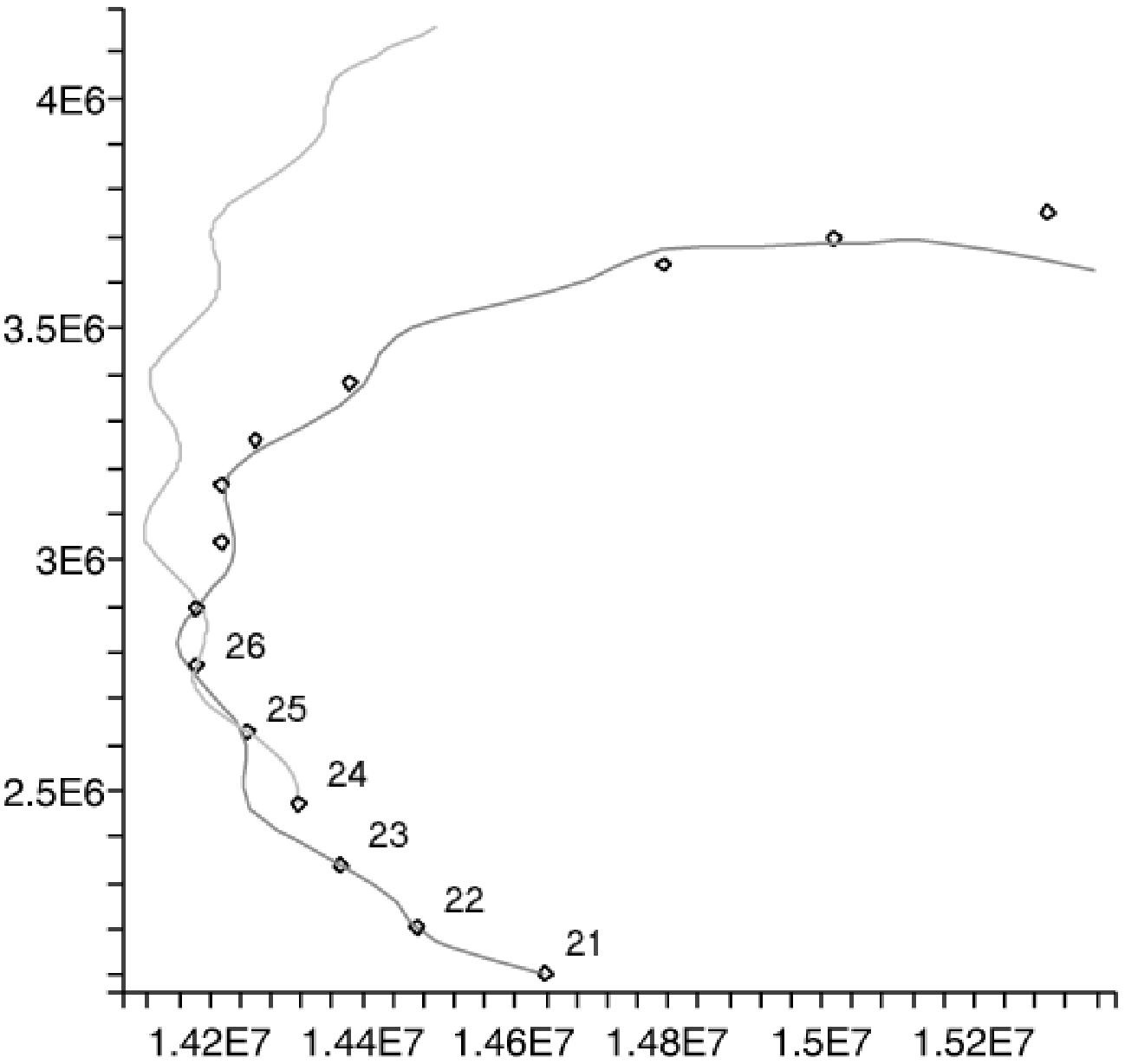}}
\caption{$\varepsilon=3\cdot 10^{-6}$: two best
predictions}\label{fig13}
\end{minipage}
\end{figure}

If we set $\varepsilon=2\cdot 10^{-6},$ we get only two appropriate
three point sets: one of them begins from the 21th point and gives a
very good prediction (Fig.\ref{fig11}), here $b_0\approx 1\cdot
10^{-5} \, s^{-5}$). The other one predicts a loop beginning from
the point 7 (see Fig.\ref{fig12}). Thus, one can assume that at the
9th point, the vortex had gone through an exterior force. It
constrains the trajectory to change the direction of its path.

If we weaken the requirement for accuracy and set
$\varepsilon=3\cdot 10^{-6},$ we get one more three point set,
beginning from point 24. In the latter case, the artificial
trajectory gives a true direction. But comparing with the exact
track, the artificial trajectory outgoing from the 21th point
(Fig.\ref{fig13}) is bad. However, the spread of positions is inside
of the annual mean diameter of the 33 percent strike probability
region for 36 hours \cite{Weber}.

\subsection{Parma}
\begin{figure}
\centerline{\includegraphics[width=0.5\columnwidth]{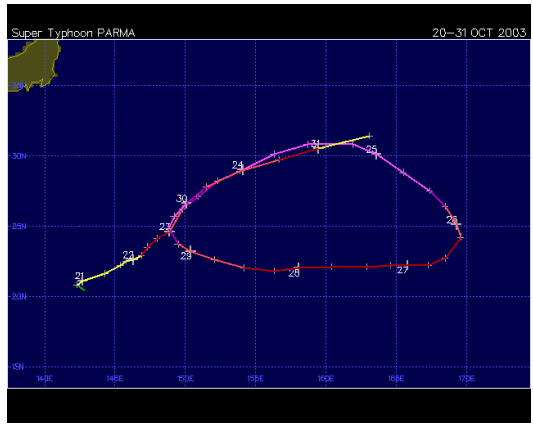}}
\caption{The Parma trajectory \cite{data} }\label{fig14}
\end{figure}

\begin{figure}
\centerline{\includegraphics[width=0.8\columnwidth]{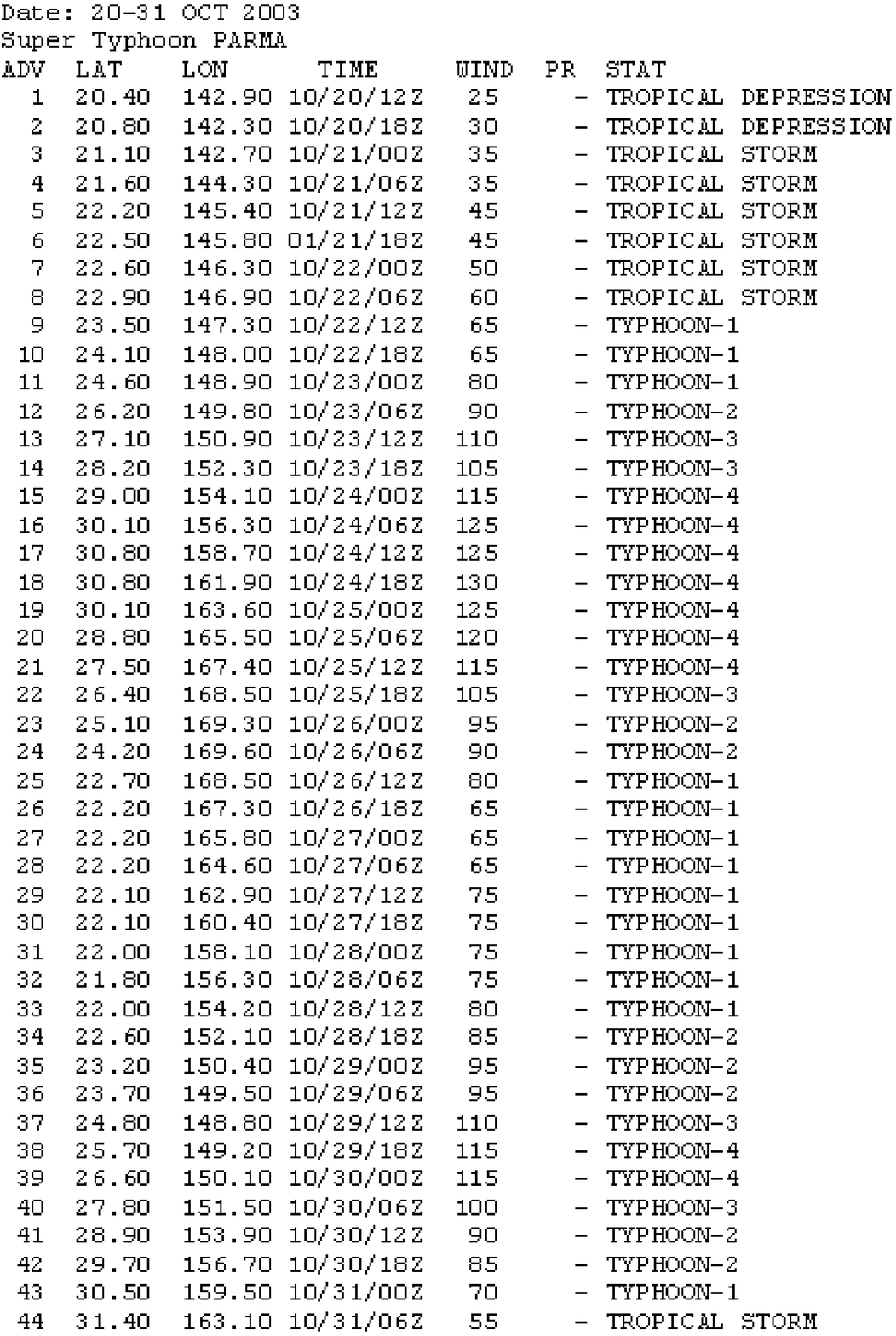}}
\caption{The Parma data \cite{data}}\label{fig15}
\end{figure}

Now we consider the example of typhoon with a looping trajectory,
Parma, 20-31 October, 2003, Fig.\ref{fig14}, Fig.\ref{fig15}.

\begin{figure}[h]
\begin{minipage}{0.5\columnwidth}
\centerline{\includegraphics[width=0.7\columnwidth]{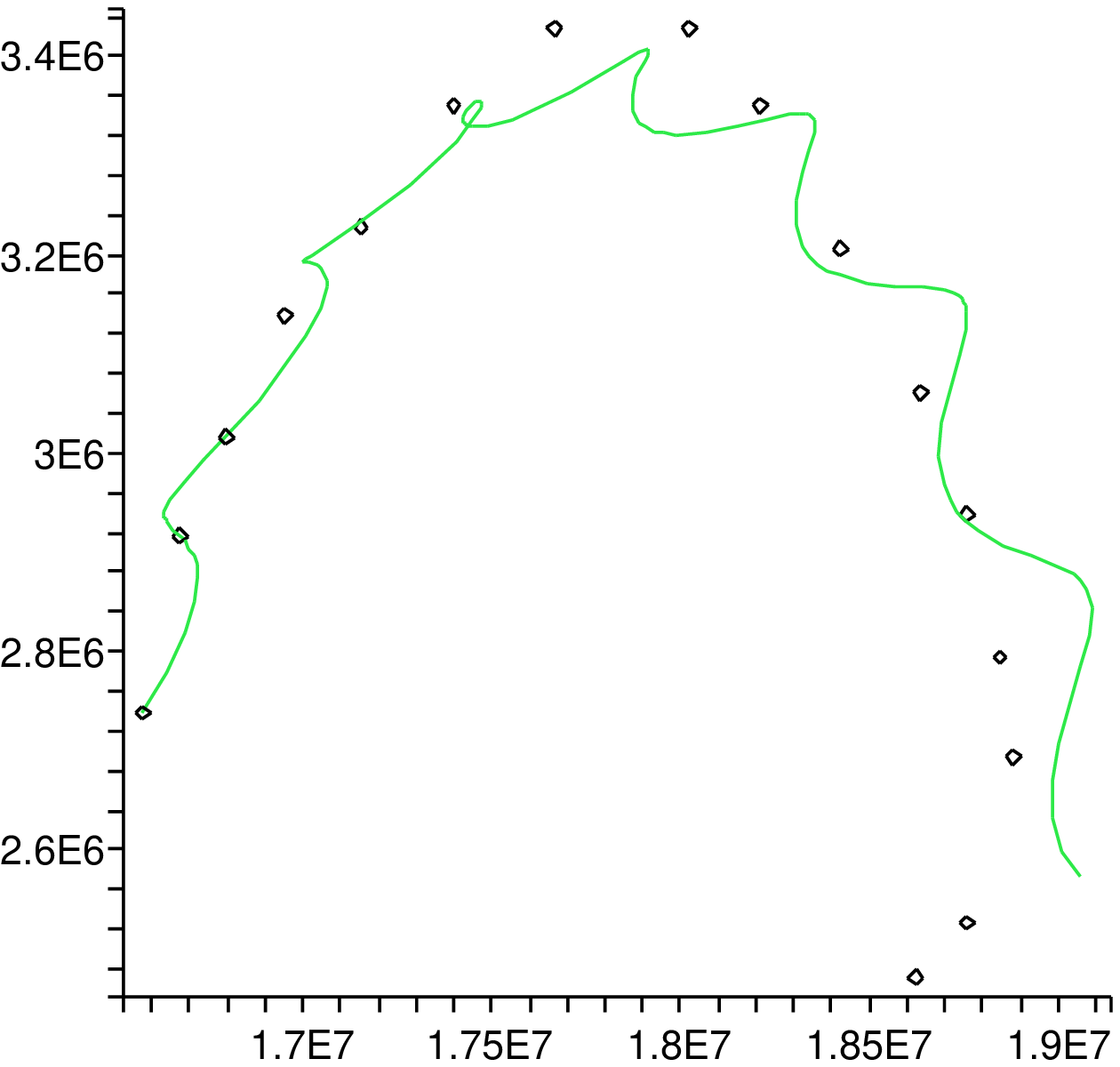}}
\caption{The Parma trajectory from \,\,\,\,\, 
 10 to 25th point (78 hours)
}\label{fig16}
\end{minipage}%
\begin{minipage}{0.5\columnwidth}
\centerline{\includegraphics[width=0.7\columnwidth]{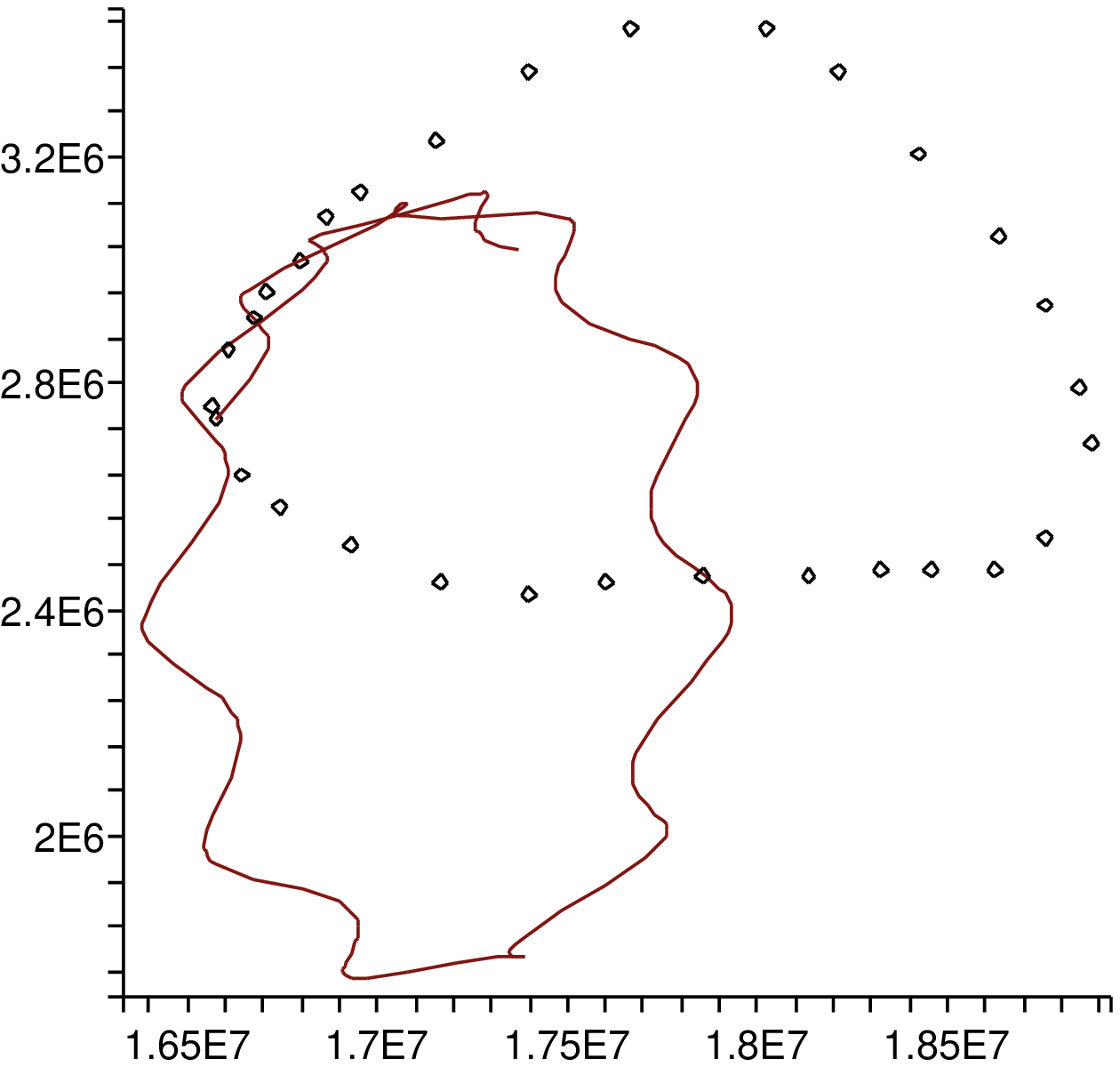}}
\caption{Forecast for 144 hours based on 10-12 points (12
hours))}\label{fig17}
\end{minipage}
\end{figure}

Fig. \ref{fig16} presents the comparison of real trajectory within
78 hours of the typhoon in its conservative phase. The parameter
$b_0=-6.1116738183 \cdot 10^{-5} s^{-1}$ is fitted for all points of
trajectory according the historical data. Fig. \ref{fig17} presents
a forecast for a long period (144 h.) based on the data for 12
hours. We can see that the forecast gives the true qualitative
behavior of the trajectory, the loop. This loop was deformed by a
steering flow.

\vskip0.7cm
\section{The role of surface friction}

It is well known that the typhoons basically do not last for a long
time over a dry land. The key point is  the significant increasing
of the dry friction when the typhoon goes to the land. Now, let us
add the damping term $-k {\bf U}$ in the right hand side of first
equation (2.8), where $k$ is a nonnegative function of coordinates.
For simplicity, we assume that $k$ is a constant. Therefore instead
of (\ref{3.51}),(\ref{3.61}), we get
\begin{equation}\dot a+a^2-b^2+lb+2c_0 A=-k a,\label{3.18}\end{equation}
\begin{equation}\dot b+2ab-la=-k b,\label{3.19}\end{equation} equations (\ref{3.41}),(\ref{3.7}) --
(\ref{3.11}) do not change.

System (\ref{3.41}), (\ref{3.18}), (\ref{3.19}) is closed. However,
it does not encounter any equilibrium for $A\ne 0.$ Therefore, it is
 not possbile to find a stable domain of low pressure from this system. Fig.18 and 19
demonstrate the break-up of the stable equilibrium for $ k > 0.$
Computer simulations are made for initial data $A_0 = 10^{-9},$
$a(0) = 0,$ $b(0) = -2\cdot 10^{-6},$ $l = 10^{-4},$ $c_0 = 0.1$ (in
respective units). Fig. 18 presents the phase portrait of the system
(\ref{3.41}), (\ref{3.18}), and (\ref{3.19}) for $k = 0.$ In this
case, a stable equilibrium exists on the phase plane. Fig. 19 shows
the collapse of vortex for $ k > 0 :$ the vortex strehgth slightly
intensifies, whereas the motion becomes significantly convergent
eventually. Here, initial data are the same as in Fig.18, $k=3\cdot
10^{-5}.$


\begin{figure}[h]
\begin{minipage}{0.5\columnwidth}
\centerline{\includegraphics[width=0.7\columnwidth]{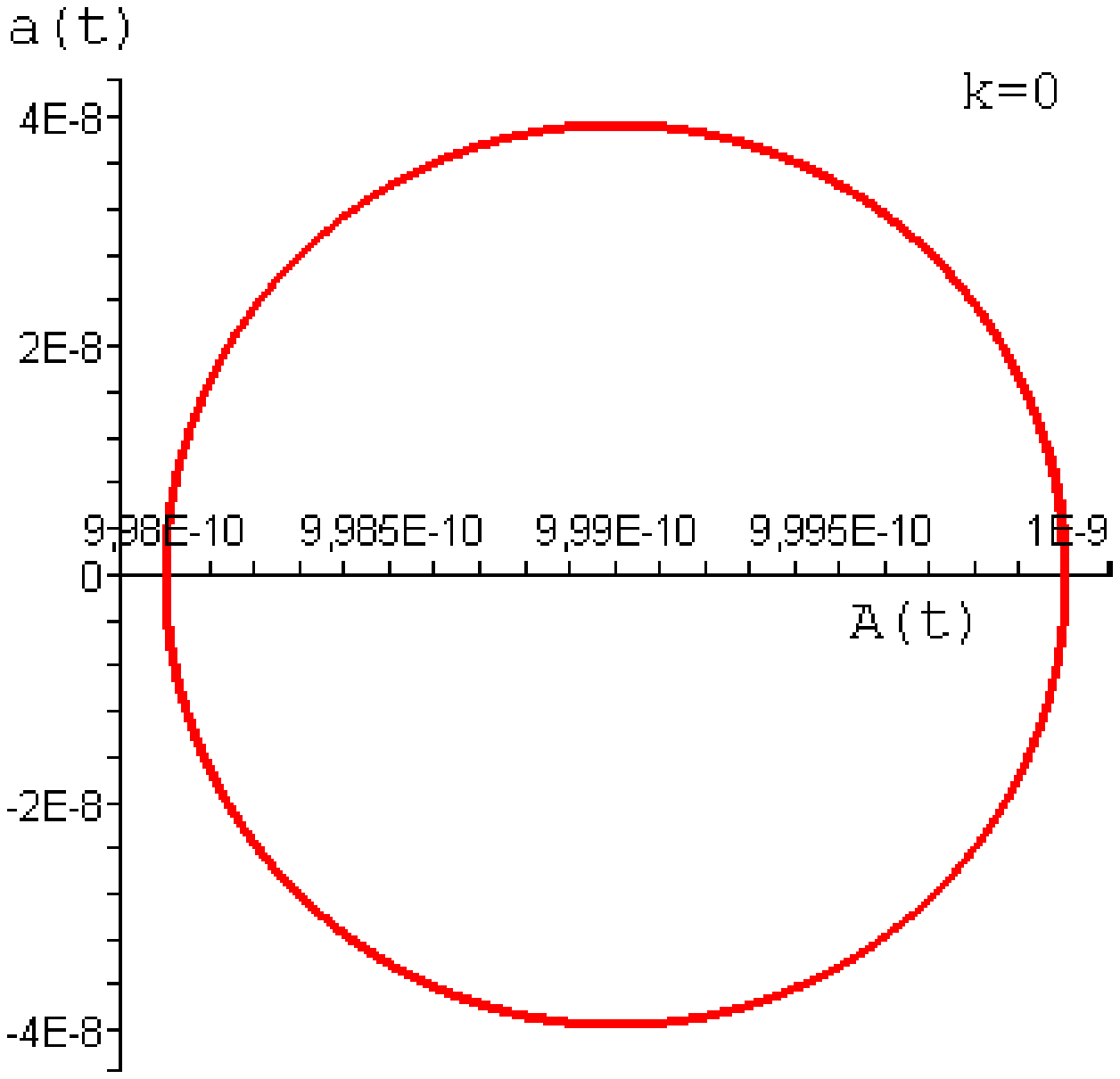}}
\caption{ The phase portrait of  (\ref{3.41}),\,\,\,\,\,\,
(\ref{3.18}),(\ref{3.19}) for $k=0.$  }\label{fig18}
\end{minipage}%
\begin{minipage}{0.5\columnwidth}
\centerline{\includegraphics[width=0.7\columnwidth]{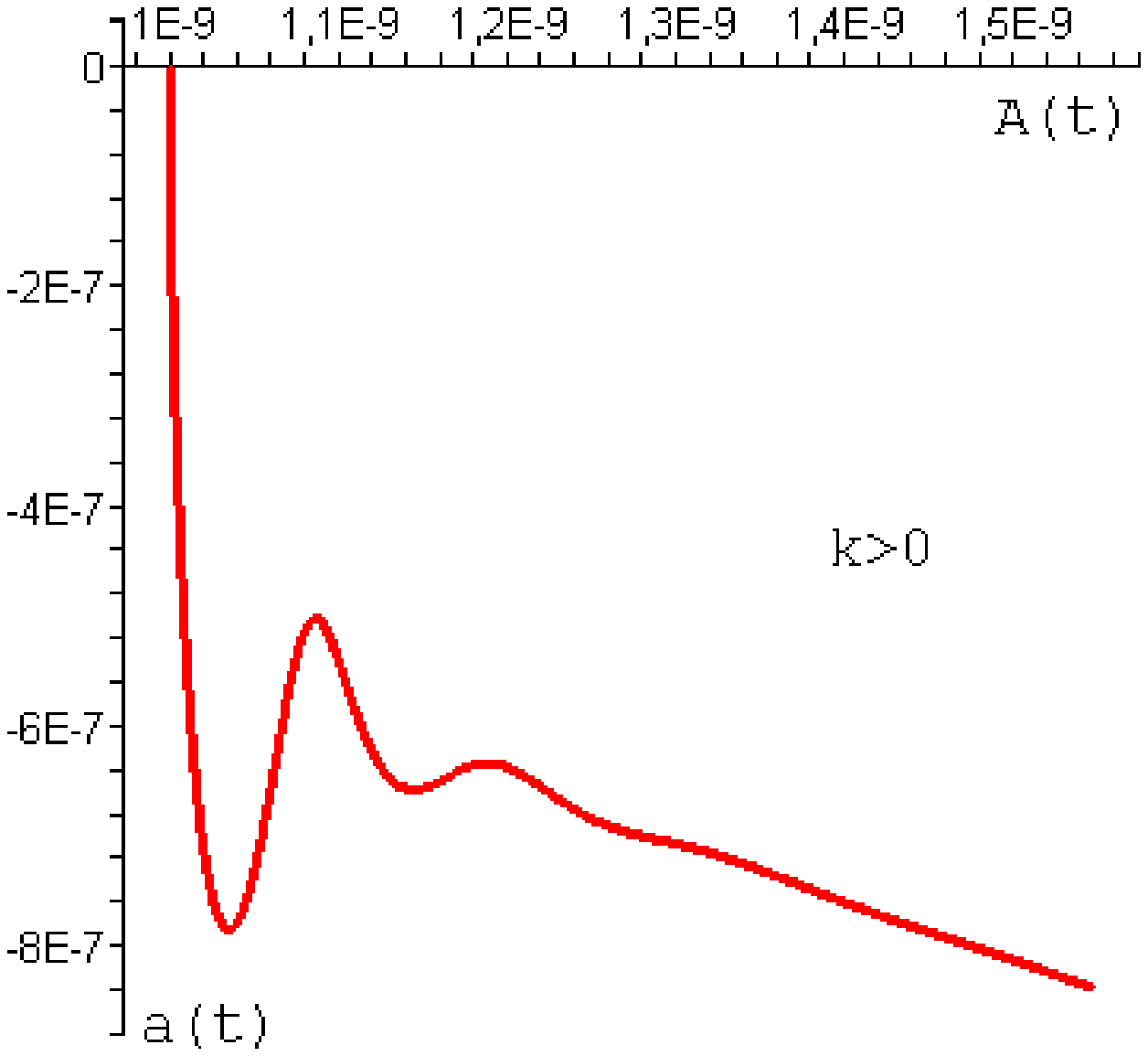}}
\caption{The same an in Fig.18 for $k=3\cdot 10^{-5},$ computations
are made for 3 days.}\label{fig19}
\end{minipage}
\end{figure}%
\vskip0.7cm


Thus, if we wish to obtain a stable vortex along with the surface
friction, we have to consider some additional phenomenon such as
vertical advection (see \cite{RFY}) in this context).

Note that the system (\ref{3.41}), (\ref{3.18}), (\ref{3.19}) is
very interesting from the mathematical point of view. It has a
complex equilibrium at the origin and there is a possibility of
existence of a strange attractor for certain values of parameters.




\section{Discussion}

In this paper, we show that the behavior of the developed tropical
cyclone trajectory is determinated by exterior and inner parameters;
the trajectory is a superposition of two circular motions: one has
period $2\pi/l,$ the other one has period $2\pi/b_0.$

Our previous conclusion that the trajectories for both barotropic
and baroclinic models are governed by the same system of
coefficient-equations seems conflict with the fact that
baroclinicity has a strong relation with the cyclone formation.
About this point, we would like to clarify that the cyclone
formation in our paper only refers to the conservative (or near
conservative) phase of the atmosphere vortex development. Indeed,
the evidence about the velocity in a linear form (\ref{3.1}) can be
confirmed by the observational data which is collected as the
developed typhoon is close to its center. As it shows in (Fig.1), at
the stage of the vortex formation, the velocity may not to be
linear. Therefore, we claim that we find a ``toy" solution for a
very complicated system to describe the processes of formation and
decay of cyclones for which an analytical solution is barely to be
obtained. It worth to remark again that the term ``barotropicity" in
our paper is only restricted to describe the bidimensional
framework. i.e. the term ``barotropicity" is used with a hidden
meaning of ``taking an average over the height". In
the physical 3D space, the flow does not have to be barotropic. \\

Our arguments can be verified from other researchers, for examples,
in \cite{XuXie}, it says ``The curvature of storm track is
determined by a set of `controlling parameters' related to the storm
characteristics and ambient atmospheric circulation. These
controlling parameters include the speed of storm (${\bf V}(0)$ in
our notation), storm intensity ($b(0)$), size ($A(0)$), the Coriolis
parameter ($l$), ambient atmospheric pressure field ($M(0), N(0)$)
and surface friction ($k$)". Also in \cite{chan}, it says that ``In
a barotropic framework, a tropical cyclone is basically ``steered"
by the surrounding flow ($V(0), M(0), N(0)$), but its movement is
modified by the Coriolis force and the horizontal vorticity gradient
of the surrounding flow."\\

Certainly, it would be naive to claim that our model provides a
better weather forecast for the typhoon trajectory than those modern
models using numerical simulations. Firstly, the formula is derived
under the assumption that the vortex is at stable phase, but, this
assumption hardy holds for the real weather forecast. Therefore, we
only can deal with oscillations near the equilibrium point at most
in our model. Secondly, the only variables that we feed in the
equations of thermodynamic parameters are initial data from real
cases. However, as the vortex moves, it may experience a forcing of
baric fields, which may skew its trajectory significantly. Thus,
generally, we can only expect to get the reproduction of the
trajectories qualitatively, i.e., to predict a turning of the track
without indicating the exact position of the vortex. Finally, we do
not take into consideration the curvilinear geometry of the Earth
surface. Therefore, only in a vicinity of an initial point, the
linearization does not result in a big distortion. The inclusion of
the $\beta$ - effect does not give an exact solution with respect to
the stable vortex (for approximative
solutions, see \cite{BVDD2}).\\

However, as in \cite{Weber}, even high-quality weather forecast
models may not be reliable due to its uncertainty about the initial
conditions or for some other unknown reasons. Thus, operational
weather forecasters still need to judge whether or not the
prediction results should be taken into consideration to build an
official weather forecast in their daily work. Therefore, it is
helpful to have a complimentary tool to aid other existing models
for the typhoon trajectory prediction. Our model seems very simple
comparing with models used for the numerical weather forecast.
However, it is very useful since it can be used to explain the
trajectory behavior, predict the direction of the trajectory and the loop formation as \vspace{0.8 cm} well.\\

Our models can be further refined in two aspects: doing analytical
refinement of the model and searching a better way for
parameters-fitting. We conjecture that it is possible to obtain some
solutions with analytical format for the Navier-Stokes system in
spherical coordinates. Then we could include the $\beta$ - drift in
our model and release the requirement that the solutions need to be
bound around the neighborhood of the center of typhoon. The natural
question arising in the parameters-fitting method is: can we
actually use the average vorticity $b_0$ as a measurement of
predictability in the model? As it is mentioned in the previous
sections, even feeding with the real data for the average vorticity
$b_0,$ one can only get approximate predication for the trajectory
behavior. The quality of the predication strongly depends on the
ambient meteorological fields, which are responsible for the
steering effect. In particular, the``regularity" of the ambient
meteorological fields seems to be a very crucial factor for the
trajectory tracing. However, it can not be obtained only through
analyzing the equations of trajectories. Thus, to improve the
predictability and accuracy for our model, we need to combine our
method with the analysis of available meteorological data and
include other theorems like statistical moments (see in this context
\cite{DOR}) and neural networks \cite{Tirozzi1}, \cite{RT} in our
model.

\begin{center}{ACKNOWLEDGMENTS}\end{center}
This work was supported by the National Science Council of Taiwan
under Grand Nos. NSC 96-2911-M001-003-MY3 and NSC
96-2112-M017-001-MY3 and National Center for Theoretical Sciences in
Taiwan. O.Rozanova was also supported by the special program of the
Ministry of Education of the Russian Federation "The development of
scientific potential of the Higher School", project 2.1.1/1399.

\end{document}